\documentclass[noshowpacs,prl,aps,superscriptaddress,floatfix,twocolumn,nofootinbib]{revtex4-2} 
\usepackage[utf8]{inputenc}
\usepackage{lineno,mathtools,url}
\usepackage{amsmath}
\usepackage{soul}
\usepackage{amssymb,mathrsfs}
\usepackage{physics}
\usepackage{graphicx}
\usepackage{hyperref}

\hypersetup{
     colorlinks=true,
     linkcolor=blue,
     filecolor=blue,
     citecolor = orange,      
   }
\usepackage[dvipsnames]{xcolor}
\usepackage{dsfont}
\usepackage{mathbbol} 
\usepackage[normalem]{ulem}
\allowdisplaybreaks
\begin{document}

\title{Superfluid Fraction and Leggett Bound in a Density Modulated  Strongly Interacting Fermi Gas at Zero Temperature}

\author{G. Orso }
\email{giuliano.orso@u-paris.fr}
\affiliation{Universit\'e Paris Cit\'e, Laboratoire Mat\'eeriaux et Ph\'eenom\`enes Quantiques (MPQ), CNRS, F-75013, Paris, France}
\author{S. Stringari}
\email{sandro.stringari@unitn.it}
\affiliation{Pitaevskii BEC Center, CNR-INO and Dipartimento di Fisica, Universit\`a di Trento, Via Sommarive 14, 38123 Povo, Trento, Italy}

\date{\today}

\begin{abstract}

We  calculate the superfluid fraction of  an interacting Fermi gas, in the presence of a one-dimensional periodic potential of strength $V_0$ and wave-vector $q$. Special focus is given to the unitary Fermi gas, characterized by the divergent behavior of the s-wave scattering length.    Comparison with the Leggett's upper bound 
$(\langle n_{1D}\rangle <1/n_{1D}>)^{-1}$, with $n_{1D}$ the 1D column density,  explicitly shows that,  differently from the case of a dilute interacting Bose gas, the bound significantly overestimates the value of the superfluid fraction, except in the phonon regime of small $q$. Sum rule arguments show that the combined knowledge of the Leggett bound and of the actual value of the superfluid fraction allows for the determination of curvature effects providing the deviation  of the dispersion of the Anderson-Bogoliubov mode from the linear phonon dependence. The comparison with the predictions of the weakly interacting BCS Fermi gas points out the crucial role of two-body interactions.
The implications of our predictions on the anisotropic behavior of the sound velocity are also discussed.
\end{abstract}

\maketitle

\section{I. Introduction}

In 1970 Leggett \cite{Leggett1970} 
derived an upper bound to the superfluid fraction  $f_{S}=n_S/\bar{n}$, given by the ratio between the superfluid density relative to the motion of the fluid along the $x$-direction and the mean density,  in terms of the spatial average of  the inverse 
of the 1D column density $n_{1D}(x)=\int n(\mathbf r) dy dz$ \cite{notefS}: 
\begin{equation}
f_{S}\le  (\langle {n}_{1D}(x)\rangle \langle \frac{1}{n_{1D}(x)}\rangle)^{-1} \equiv f^L_{S} . \; 
\label{LB}
\end{equation}
Result (\ref{LB}) is particularly relevant at zero temperature, where the superfluid fraction is affected by the density modulations induced by the breaking of translational invariance, and holds independent of whether the breaking occurs spontaneously  or it is induced by an external force. At finite temperature the superfluid fraction is instead affected by thermal  effects even for uniform systems, with important consequences  on the propagation of second sound \cite{Peshkov:JPhys1944,Sidorenkov:Nature2013,Christodoulou:Nature2021}.
Under the assumption that the many-body wave function is  described by a single phase  equal for all the particles,  an assumption applicable to a dilute Bose gas at zero temperature described by the Gross-Pitaevskii equation, Leggett derived a further lower bound \cite{Leggett1998} to the superfluid fraction. The Leggett upper and lower bounds coincide if the 3D density depends only on the $x$ variable, so that in this case the inequality (\ref{LB}) reduces to an equality.
   In a recent experiment \cite{Chauveau:PRL2023} the Leggett's integral entering Eq.(\ref{LB}) was actually measured in an ultracold Bose gas confined in a 2D box, in the presence of a periodically modulated external 1D potential, causing important disomogeneities in the density profile, and the bound (\ref{LB}) was shown to agree with good accuracy with the result for the superfluid fraction measured through the velocity of sound, in agreement with the predictions of Gross-Pitaevskii theory. 

The question of the accuracy of Leggett  bound  in other interacting many-body systems is basically unexplored 
and the aim of our work is to address the question in the case of an interacting Fermi gas  at zero temperature, with special focus on the so-called unitary limit, where the $s$-wave scattering length is infinite and the system behaves as a strongly interacting
superfluid.  We will consider a gas confined in a box  in the presence of a periodic potential of the form
\begin{equation}
V_{ext}= V_0 \cos(qx)
\label{Vext}
\end{equation}
which induces density modulation of period $d=2\pi/q$ in the ground state density profile. 
We will  attack the problem following two different lines. In Sect. II we develop a sum rule approach to explore the behavior of the superfluid fraction in the presence of a weak perturbation (small $V_0$)  and calculate the  Leggett's integral as well as  the exact value of the superfluid fraction. The sum rule results  provide the proof that  only if the excitation spectrum of the unperturbed uniform superfluid consists of a single excitation exhausting the relevant sum rules of the dynamic structure factor, will the inequality (\ref{LB}) reduces to an identity. This is the case of the dilute BEC gas for all  values of the wave vector $q$,   but remains an inequality in the most general case. The results of the sum rule approach are also used to infer on the curvature of the 
 dispersion of the Anderson-Bogoliubov collective mode of a Fermi superfluid at long wavelengths. 
In Sect. III we  first provide a quantitative implementation of the sum rule approach developed in Sect II by applying the dynamical Bardeen–Cooper–Schrieffer (BCS) theory \cite{Combescot:PRA2006}. We then  present our numerical results for the superfluid fraction of the unitary Fermi gas by solving the Bogoliubov-de Gennes (BdG) equations in the presence of the external  periodic potential (\ref{Vext})  in order to provide a quantitative comparison between the Leggett bound  and the actual value of the superfluid fraction, beyond the weak perturbation limit of small $V_0$,  in a wider range of parameters of possible relevance for future experiments.
In Sect IV we discuss how our predictions for the superfluid density could be tested by measuring the effects of anisotropy in the velocity of sound.

\section{II. Sum rule approach}
At zero temperature, the dynamic structure factor $S(\omega, q)$ of a many-body system  is related to the imaginary part of the density response function $\chi(\omega,q)$ by 
\begin{equation}\label{SqImchi}
S(\omega, \mathbf{q})=-\frac{1}{\pi}\textrm{Im}\chi(\omega+i0^+, \mathbf{q}),    
\end{equation} 
so that its spectral representation is
\begin{equation}\label{Swq}
S(\omega, \mathbf{q})=\frac{1}{V}\sum_{n}|\langle 0|\rho_\mathbf q |n\rangle|^2 \delta(\omega+E_0-E_n),    
\end{equation}
where $V$ is the volume of the box, $\rho_\mathbf q= \sum_k \exp(i \mathbf q \cdot \mathbf x_k)$ is the Fourier transform of the density operator, $|n\rangle$ is a complete set of eigenstates of the  many-body Hamiltonian and $(E_n-E_0)$ are the corresponding excitation energies. Here and in the following we use 
the convention $\hbar=1$ and consider configurations where the dynamic structure factor $S(\omega, \mathbf{q})$
 depends on the  wave-vector only through its modulus $q=|\mathbf q|$. From the dynamic structure factor  one can extract the energy weighted moments 
\begin{equation}\label{moments}
m_j(q)=\frac{1}{\bar n}\int_0^\infty d\omega \omega^j S(\omega,q)    
\end{equation}
where $j$ is an integer and $\bar n=N/V$ is the average fermion density.  The energy weighted moment $m_1(q)$ is model independent and equal to $q^2/(2m)$, as follows from the number conservation law (f-sum rule), with $m$  the particle mass. All the other moments are generally model dependent and do carry crucial information on the many-body properties of the system, as we shall see below.

\subsection{A. Superfluid fraction $f_S$ vs Leggett bound $f_S^L$}

If the periodic potential (\ref{Vext}) is weak (small $V_0$), the density changes induced by the perturbation can be calculated via linear response theory and take the form

\begin{equation}\label{dn}
\delta n(x) = V_0 \chi(0,q)\cos(qx) \; ,\\
\end{equation}
where $\chi(\omega=0,q)$ is the static  response function,
related to the inverse energy weighted moment of  the dynamic structure factor  $S(\omega,q)$ by \cite{pines1966theory}:
\begin{equation} \label{chi0}
-\frac{1}{\bar n}\chi(0,q) =2 m_{-1}(q) 
\end{equation}  \;
By expanding the Leggett's integral up to terms quadratic in $V_0$, one finds the following result for the Leggett's upper bound:
\begin{equation}
f^L_S= 1-2V^2_0(m_{-1}(q))^2
\label{fsL-1}
\end{equation}
in terms of $V_0$ and $m_{-1}(q)$. Since Eq.(\ref{fsL-1}) is based on linear response theory, the moment $m_{-1}(q)$ 
in Eq.(\ref{fsL-1}) has to be evaluated using the unperturbed ($V_0=0$) uniform system.  

In the following we will consider many body systems, characterized by the absence of  transverse gapless excitations. This includes, among others, single component Bose and Fermi superfluids in the presence of an  external periodic perturbation  of the form (\ref{Vext}). In order to calculate the superfluid fraction along the $x$-direction it is convenient to calculate the expectation value $\langle P_x\rangle$ of the longitudinal momentum $P_x=\sum_{k=1}^Np_{k,x}$ acquired by the system in the presence of the longitudinal perturbation $-v P_x$  and use the definition
\begin{equation}
f_S= 1 - \lim_{v\to 0} \frac{\langle P_x\rangle}{Nmv}
\label{fS}
\end{equation}
By applying perturbation theory  the superfluid fraction can then be written in the form 
\begin{equation}\label{pert}
f_S=1 - \frac{2}{Nm}\sum_{n\ne 0}\frac {|\langle 0 |P_x |n\rangle|^2}{E_n-E_0}.
\end{equation}
If the Hamiltonian is translationally invariant, i.e. if $[H,P_x]=0$, then $f_S=1$. If instead the Hamitonian contains the additional term (\ref{Vext}) breaking translational invariance,  then the system will acquire a normal component. Using the commutation relation
\begin{equation}
[H, P_x]= -\frac{q}{2}V_0(\rho_q-\rho_{-q})  \; ,
\end{equation}
one can rewrite Eq.(\ref{pert}) in the form
\begin{equation}
f_S=1 -\frac{1}{Nm}V_0^2q^2\sum_{n\ne 0}\frac {|\langle 0|\rho_q |n\rangle|^2}{(E_n-E_0)^3}= 1-2 V_0^2m_1(q)m_{-3}(q),
\label{fsm-3}
\end{equation}
showing that the superfluid fraction is controlled by the inverse cubic energy weighted moment $m_{-3}(q)$.
While Eq.(\ref{fsL-1}) is  valid only for a weak periodic potential, Eq. (\ref{fsm-3}) holds for arbitrary values of $V_0$. Of course for small $V_0$ the moment $m_{-3}$ can be safely evaluated using the unperturbed uniform system. 

Eq.s (\ref{fsL-1}) and (\ref{fsm-3}) are the main results of this section. They show that, in the considered limit of weak  perturbations (small $V_0$), the Leggett's bound coincides with the exact value of the superfluid fraction only if the general inequality
\begin{equation}
\frac{m_{1}}{m_{-1}}\ge \frac{m_{-1}}{m_{-3}} \; 
\label{inequality},
\end{equation}
which follows from the positiveness of the domain of the dynamic structure factor, reduces to an equality.  The equality holds  only in few notable cases:
i) for a dilute BEC gas, where the Bogoliubov excitation spectrum is characterized by a single mode for all values of the wave vector $q$, with frequency  $\omega_=\sqrt{q^2 \mu_b/m + q^4/(2m)^2}$, where $\mu_b$ is the mean field chemical 
potential of the BEC gas.
ii) In general for superfluids in the phonon limit $q\to 0$, as a consequence of the fact that in this limit the phonon mode exhausts the relevant sum rules entering the  inequality (\ref{inequality}). This includes, in particular the case of the  interacting superfluid Fermi gas considered in this paper but also the case of  superfluid $^4He$. 
In general the inequality (\ref{inequality}) implies that $f_S^L \ge f_S$, in agreement with Leggett's prescription. 

It is useful to stress that the existence of a finite $m_{-3}$
for a Fermi gas requires the particle-hole excitations spectrum to be gapped (as it is indeed the case in the superfluid state), 
otherwise $m_{-3}$ diverges  due to the presence of gapless particle-hole excitations at the Fermi surface \cite{pines1966theory}.

\subsection{B. Curvature effect in the  excitation spectrum}

Interestingly, the moments $m_{-1}$ and $m_{-3}$, controlling the behavior of the 
Leggett bound $f_S^L$ and of  the actual value $f_S$ of the superfluid fraction in the presence of the periodic potential (\ref{Vext}),
can also be used to characterize  the dispersion relation of the Anderson-Bogoliubov mode of a uniform system  at small $q$
\begin{equation}
\omega^2 = c^2q^2(1+\gamma q^2),
\label{expansion}
\end{equation}
where $c$ is the  sound velocity, related to the inverse compressibility $\kappa^{-1}=n (\partial \mu/\partial n)$, through the thermodynamic relation $c^2=\kappa^{-1}/m$,
and $\gamma$ measures the curvature effect beyond the linear dispersion.
The coefficient $\gamma$ is positive in the case of a dilute BEC gas $(\gamma =1/(4\mu_b)$),   while it is expected to become negative in the BCS region of the BEC-BCS crossover. Its change of sign takes place around unitarity  and has been the object of several theoretical calculations \cite{SON2006197,Haussmann:PRA2009,Kurkjian:PRA2016,Zou:PRA2018} and of a recent measurement in a superfluid atomic Fermi gas using  Bragg spectroscopy \cite{Biss:PRL2022}. 

To understand the connection between the results for the superfluid fraction in the presence of the perturbation $V_0\cos(qx)$ and the curvature effect, it is useful to explore the behavior of  
the contributions to the various sum rules originating from the gapless branch and from the gapped excitations in the small $q$ regime (see Table 1).  The $q^4$ dependence of the strenght contribution of the gapped states  follows from general arguments based on Galilean invariance (see for example \cite{pines1966theory}) and implies  that the $m_{-3}$ and $m_{-1}$ moments of the dynamic structure factor  are exhausted  by the gapless mode not only  in the lowest $q$ phonon regime,  but also including the  first corrections in $q^2$. 
\begin{table}
\caption{\label{tab:table1}%
Low momentum behavior of the leading contributions 
 to a given physical quantity (left column),
 coming from gapless collective modes (central column)  and from gapped  single-particle excitations (right column). Notice that the  sum of  the $q^4$ contributions of the gapless and gapped excitations to the energy weighted moment $m_1$ cancel out in order to ensure the fulfillment of the $f$-sum rule,   $m_1(q)=q^2/(2m)$. }
\begin{ruledtabular}
\begin{tabular}{lcr}
& Gapless & Gapped \\
$|\langle0|\rho_q|n\rangle|^2$ & $q + q^3$ & $q^4$ \\
$\omega(q)$ & $q+q^3$ & const\\
$m_1(q)$ & $q^2+q^4$ & $q^4$ \\
$m_{-1}(q)$ & const + $q^2$ & $q^4$\\
$m_{-3}(q)$ & $1/q^2$ + const & $q^4$ \\
\end{tabular}
\end{ruledtabular}
\end{table}
A non trivial consequence is that, at low $q$, the average square excitation energy 
\begin{equation}\label{omega2}
\omega^2=  m_{-1}(q)/m_{-3}(q)
\end{equation}
exactly coincides with the expansion (\ref{expansion})  of the square frequency, including the curvature correction in $\gamma$. Viceversa, the most popular ratio $m_1(q)/m_{-1}(q)$  correctly reproduces only the leading phonon term $c^2q^2$. 
Using the small $q$ expansions 
$m_{-1}(q)= (\kappa/2) (1-\beta_{-1}q^2)$  and $q^2m_{-3}(q)= (m\kappa^2/2)(1-\beta_{-3}q^2)$ in Eq.(\ref{omega2}) and comparing the obtained expression with Eq.(\ref{expansion})  one finds 
\begin{equation}
\gamma= \beta_{-3}-\beta_{-1}.    
\end{equation}
By substituting the same expansions in  Eq.s (\ref{fsL-1}) and (\ref{fsm-3}), we note that the inequality $f_S\leq f_S^L$ yields $\beta_{-3} \leq 2\beta_{-1}$.
When  $\gamma$ vanishes, corresponding to $\beta_{-1}=\beta_{-3}$, the $q^2$ corrections to the Leggett bound and to the exact superfluid fraction differ exactly by a factor $2$. 
In the case of a superfluid Fermi gas this happens very close to unitarity \cite{Biss:PRL2022}.  The situation is different for a dilute BEC gas, where $\beta_{-3}=2\beta_{-1}$, since $f_S^L=f_S$ for all values of $q$. 

It's also interesting to compare the above results with the effective field theory 
for the unitary Fermi gas developed in \cite{SON2006197}, which includes the 
leading order corrections in an expansion in small gradients of the 
quantum hydrodynamic Lagrangian density (see also the related discussion in Ref.\cite{Haussmann:PRA2009}). The corresponding predictions 
for the static response function and for the dispersion of the collective mode at small $q$ are given by
\cite{SON2006197}
\begin{eqnarray}
 \chi(0,q)&=&-n\kappa \left[1+A   \left(c_1-\frac{9}{2}c_2\right ) \frac{q^2}{k_F^2} \right], \label{sonchi} \\
 \omega^2 &=& c^2q^2 \left[1-A \left(c_1+\frac{3}{2}c_2\right ) \frac{q^2}{k_F^2} \right], \label{sonomega2}
\end{eqnarray}
where $c_1$ and $c_2$ are dimensionless parameters entering the Lagrangian density, 
$k_F=(3\pi^2 n)^{1/3}$ is the Fermi wave-vector, $A=2\pi^2\sqrt{2(1+\beta)}$ and $\beta$ is the dimensionless  Bertsch parameter entering the equation of state of the homogeneous Fermi gas at unitarity:  $\mu=(1+\beta) k_F^2/(2m)$.  
The BdG equations yield $\beta= -0.41$ (see Sec. III), 
while more accurate Monte Carlo calculations predict  $\beta=-0.58$ \cite{Carlson:PRL2004,Astrakharchik:PRL2004}, which is quite close to the value $\beta= -0.63$ reported in the experiment \cite{Ku:Science2012}.

By comparing Eqs. (\ref{sonchi}) and (\ref{sonomega2})
with Eq.(\ref{chi0}) and (\ref{omega2}), respectively, we find  that the 
parameters of the effective field theory can be expressed as
 \begin{eqnarray}\label{eft}
c_1&=& (2 \beta_{-1}-3\beta_{-3})\frac{k_F^2}{4A},\nonumber \\
c_2&=&(2 \beta_{-1}-\beta_{-3})\frac{k_F^2}{6A}.
 \end{eqnarray}
In Ref.\cite{Haussmann:PRA2009} it was demonstrated that the coefficient $c_2$ must be positive.  From Eq.(\ref{eft}) we see that 
this is equivalent to the condition $f_S<f_S^L$.

\section{III. Numerical results}

\subsection{A. Dynamical BCS theory and sum rules}
In this section we evaluate the sum rule moments of a unitary Fermi gas with the help of the dynamical BCS theory developed in Ref.\cite{Combescot:PRA2006} (see also an earlier work by Minguzzi and collaborators \cite{Minguzzi:EPJD2001}). 
In particular we will make use of the general expression for the  density response function 
\begin{eqnarray}
\chi(\omega, q)&=&-\frac{1}{2\pi^2} \Big( I^{\prime \prime} (\omega,q) \nonumber \\ 
&-&\frac{F(\omega,q)}{I_{11}(\omega, q)  I_{22}(\omega, q)-\omega^2 I_{12}^2(\omega,q)}\Big),   \label{defchi} 
\end{eqnarray}
where $I^{\prime \prime}, F$ and $I_{ij}$ are functions, whose definition  in terms of integrals over momenta can be found in Ref.\cite{Combescot:PRA2006}). Notice that, with respect to the
result quoted in Ref.\cite{Combescot:PRA2006} (see their Eq.(25)),
the rhs of Eq.(\ref{defchi}) contains an extra factor of 2, because in our definition of the density response function we account for the spin degeneracy [see Eq.~(\ref{dn})].   The first term in Eq.(\ref{defchi}), proportional to $I^{\prime \prime}$, represents the bare 
contribution  from the linear response formalism applied to the BCS state, while the second term accounts for the fluctuations of the order parameter over its mean-field value, which are included via the random phase approximation  (RPA). In particular the vanishing of the denominator of the second term, $ I_{11}(\omega, q)  I_{22}(\omega, q)-\omega^2 I_{12}^2(\omega,q)=0$, yields the dispersion relation of the  Anderson-Bogoliubov mode, $\omega=\omega_{AB}(q)$, which has a phonon-like behavior at low momentum. 

Although not explicitly written, all functions in Eq.(\ref{defchi}) depend also on the two mean-field parameters describing the BCS-BEC
crossover in Fermi gases, namely the energy gap $\Delta$ and the chemical potential $\mu$. 
Their values can be obtained by inverting the BCS gap equation and the number equation describing the BCS-BEC crossover for uniform systems, see for instance \cite{Combescot:PRA2006}. In particular, for a unitary Fermi gas, corresponding to $1/(k_Fa)=0$, one finds $\mu=0.591E_F$ and $\Delta=0.686 E_F$.
The dynamic structure factor can  be calculated numerically from Eq.s (\ref{SqImchi}) and (\ref{defchi}). For a given momentum
$q$, there is a threshold frequency $\omega_{tr}(q)$, where the collective mode merges with the continuum of single-particle excitations \cite{Combescot:PRA2006}
\begin{equation}
\omega_{tr}(q)=\left\{{2 \Delta\;\;\;\;\;\;\; \;\;\;\textrm{for}\; \mu > 0\; \text{and}\;q<2\sqrt{2m \mu} , 
\atop 2\sqrt{(q^2/(8m)-\mu)^2+\Delta^2} \;\;\;\text{otherwise.}}\right.
\end{equation}
For $\omega< \omega_{tr}(q)$, all the functions $I^{\prime \prime}, F$ and $I_{ij}$  in  Eq.(\ref{defchi}) are real, due to the energy gap in the single-particle energy spectrum. In this case the only contribution to the dynamic structure factor originates from the collective mode, as the vanishing of the denominator in the second term in Eq.(\ref{defchi}) causes the appearance of an imaginary part, via  the formula $1/(x+i0^+)=\textrm{P} 1/x-i \pi \delta(x)$, where $\textrm{P}$  stands for the principal part.
In contrast, for $\omega >\omega_{tr}(q)$,  the only contribution to $S(q,\omega)$ comes from the continuum of single-particle excitations, with all the functions $I^{\prime \prime}, F$ and $I_{ij}$  in  Eq.(\ref{defchi}) acquiring an imaginary part by themselves. 

\begin{figure}
\includegraphics[width=\columnwidth]{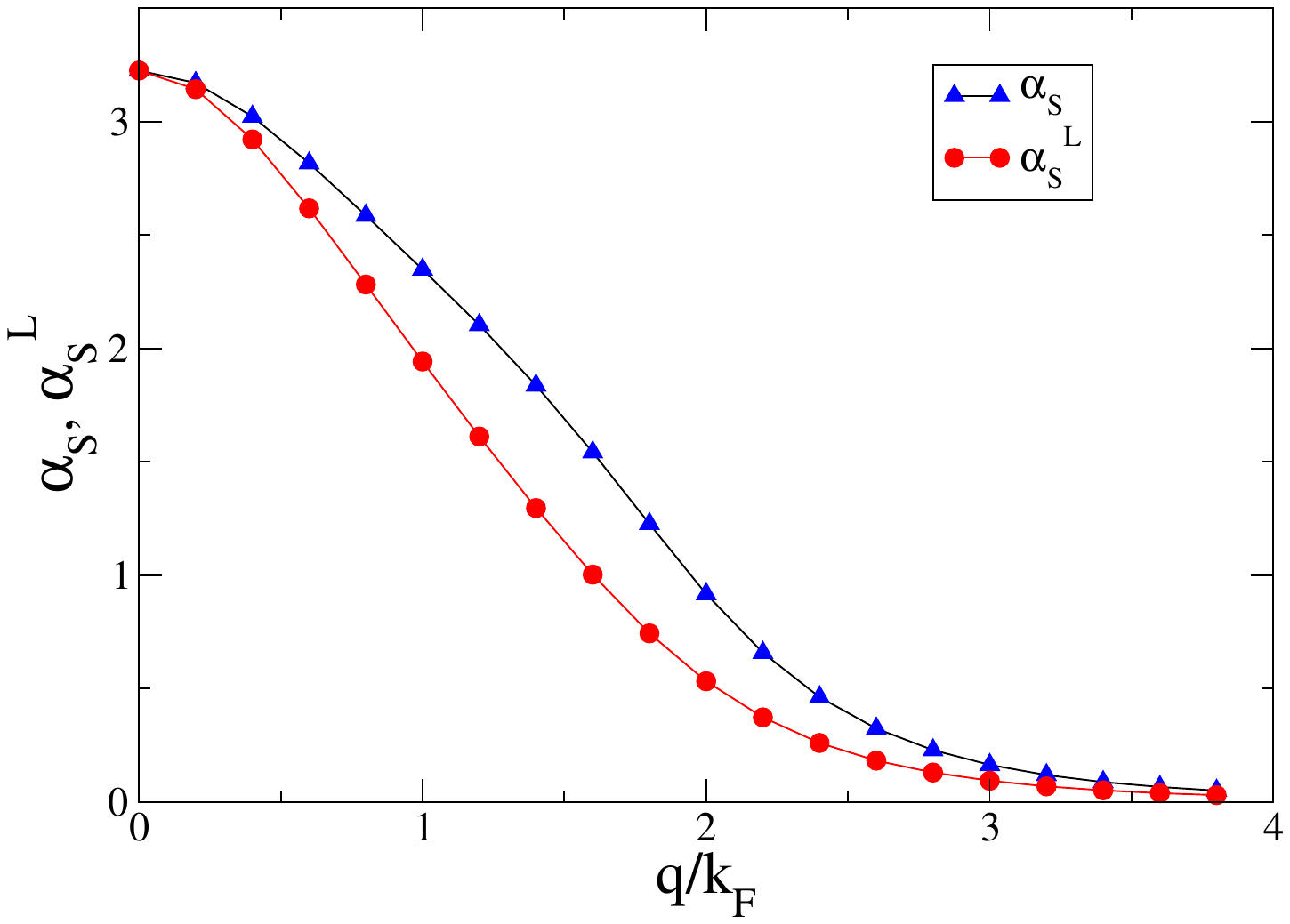}
\caption{Lattice-induced perturbative corrections to the superfluid density,  $\alpha_S=(1-f_S)E_F^2/V_0^2$,
and to the associated Leggett's bound, $\alpha_S^L=(1-f_S^L)E_F^2/V_0^2$, of a unitary Fermi gas, calculated  from the sum-rule approach (see Eq.s (\ref{fsm-3}) and (\ref{fsL-1})).
}
\label{Fig:alpha}
\end{figure}

From the dynamic structure factor we extract the moments  (\ref{moments})   and
use them to estimate the superfluid density and the associated  Leggett bound in the presence of a weak periodic potential, corresponding to $V_0\ll E_F$, with $E_F=k_F^2/(2m)$ being the Fermi energy of the unperturbed system.
Since the potential induces a quadratic in $V_0$ corrections  to both quantities, we write $f_S=1-\alpha_S (V_0/E_F)^2$ and $f_S=1-\alpha_S^L (V_0/E_F)^2$, where 
$\alpha_S$ and $\alpha_S^L$ are dimensionless functions. 
From Eq.s (\ref{fsL-1}) and (\ref{fsm-3}) we find, respectively,  $\alpha_S^L(q)=2(m_{-1}(q))^2 E_F^2$ and $\alpha_S(q)=2 m_1(q)m_{-3}(q) E_F^2$. 
In Fig.\ref{Fig:alpha} we plot such quantities as a function of the
ratio $q/k_F$ for a unitary Fermi gas. 
We see that in general  $\alpha_S^L(q)\neq \alpha_S(q)$, showing that  the Leggett bound is not exhausted in the presence of the external periodic potential. A notable exception occurs for vanishing $q$, where  the two quantities coincide and reduce to $\kappa^2 E_F^2/2=9E_F^2/(8\mu^2)=3.221$.
Moreover, we see from Fig.\ref{Fig:alpha} that the corrections to the superfluid density become less and less important as $q$ increases and vanish for infinite $q$, reflecting the behavior of the static response function.  

\begin{figure}[tb]
\includegraphics[width=\columnwidth]{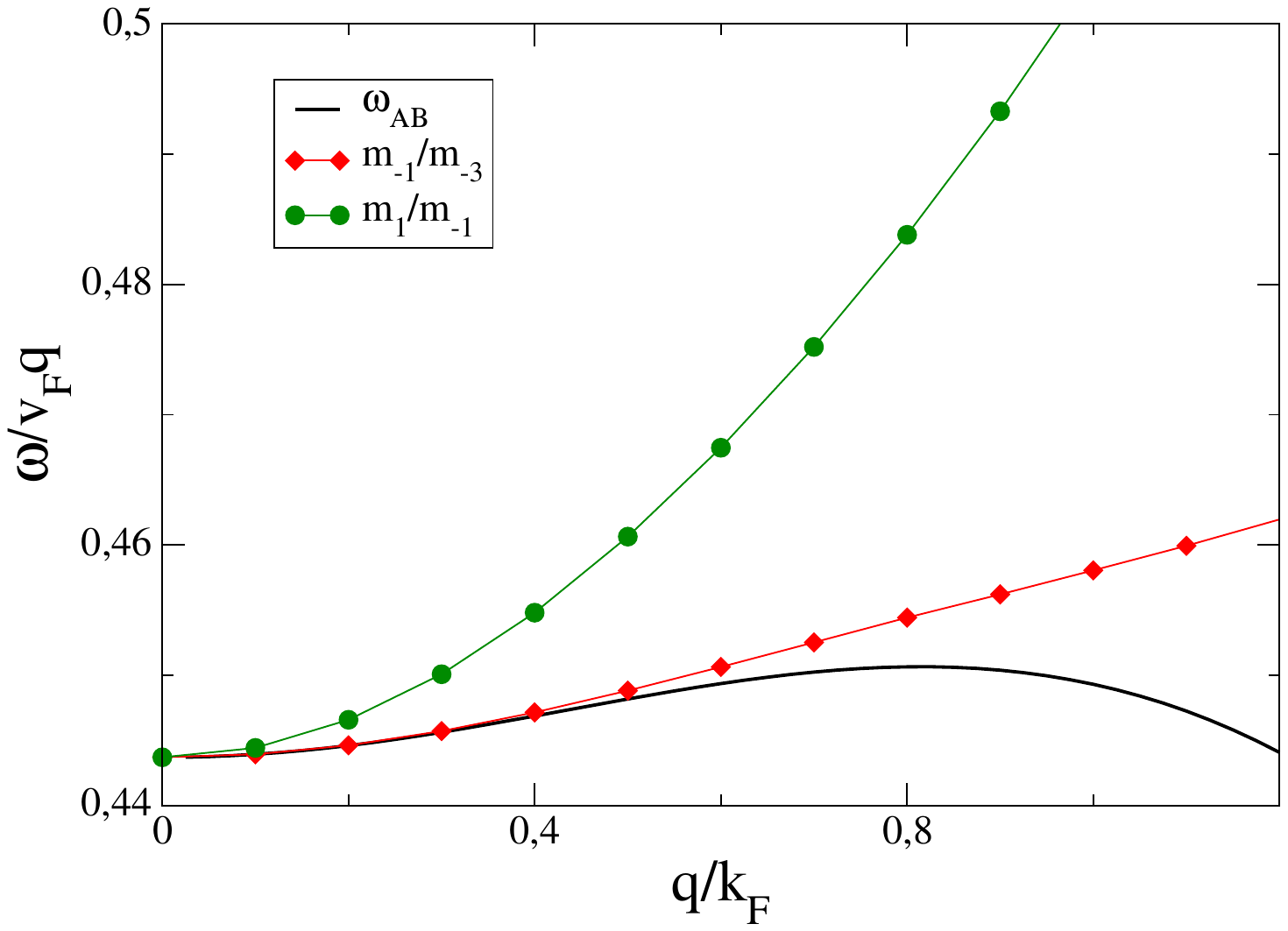}
\caption{Dispersion relation of the Anderson-Bogoliubov collective mode of the unitary Fermi gas.  The sum rule estimates (symbols) in the long wavelength limit are compared with the full numerical result $\omega=\omega_{AB}(q)$ (black solid line),  first discussed in Ref.\cite{Kurkjian:PRA2016} . Here $k_F$ is the Fermi momentum and $v_F=k_F/m$ is the Fermi velocity. The sound velocity of the system is $c=0.444 v_F$.
}
\label{Fig:disp}
\end{figure}

In Fig.\ref{Fig:disp} we plot the low-momentum estimates $\omega(q)= \sqrt{m_1(q)/m_{-1}(q)}$ (green circles) and $\omega(q)= \sqrt{m_{-1}(q)/m_{-3}(q)}$ (red diamonds) for the energy dispersion of the Anderson-Bogoliubov collective mode of a unitary Fermi gas
based on the sum rule approach.
The black solid line represents the full numerical result, $\omega=\omega_{AB}(q)$,  showing a positive, albeit small, curvature \cite{Kurkjian:PRA2016}.
At higher momentum, however,  the correction to the phononic dispersion becomes negative, as the frequency of the collective mode is bounded above by the threshold frequency $\omega_{tr}$.  
We see that the sum rule estimate $\omega(q)=\sqrt{m_1(q)/m_{-1}(q)}$  reproduces only the sound velocity of the superfluid in the phononic regime, while missing the curvature correction. In contrast, the  sum-rule estimate $\omega(q)= \sqrt{m_{-1}(q)/m_{-3}(q)}$ is more accurate as it includes the first non linear correction. 

\subsection{B.  BdG equations for systems in a periodic potential}
So far we have considered a  weak external periodic potential, so that its effect  on the transport properties of the unitary Fermi gas can be calculated within perturbation theory. Below we compute the superfluid fraction and the Leggett bound for a periodic potential of arbitrary strength, by solving numerically the BdG equations. 
Since the external potential (\ref{Vext}) is periodic, the superfluid fraction defined in Eq. (\ref{fS}) coincides with the effective mass ratio $m/m^*$ of the Fermi gas, calculated from the curvature of the lowest energy band according to
\begin{equation}\label{invm}
f_S=\frac{m}{m^*}\equiv \lim_{Q\to 0} \frac{1}{\bar{n}}\frac{\partial^2 e(\bar{n},Q)}{\partial Q^2},
\end{equation}
where $e=E/V$  is the energy density  and $Q$ is the wave-vector of the supercurrent. 
To see this, we note that -$\langle P_x \rangle$ in Eq. (\ref{fS}) represents the total momentum of the system along the x-axis  measured with respect to the moving frame, so that the supercurrent density $j$ in the lab frame is given by
 \begin{equation} \label{Eq:j}
 j=\bar n v - \frac{\langle P_x \rangle}{m V}= f_S \bar n v + O(v^2).  
 \end{equation}
Since the supercurrent density is  related to the 
wave-vector by $j=\partial e/\partial Q$, from Eq. (\ref{Eq:j})
we find $v=Q/m$
and 
\begin{equation}
e(\bar n, Q)=e(\bar n,0)+ f_S \bar n \frac{ Q^2} {2m}+ O(Q^3),  
\end{equation}
which is consistent with Eq. (\ref{invm}).
In the absence of the lattice ($V_0=0$),
one has $e(\bar{n},Q)=e(\bar n,0)+ \bar n Q^2/(2m)$,  and hence Eq.(\ref{invm})  yields $m^*=m$ and $f_S=1$.

In order to calculate the superfluid fraction via Eq. (\ref{invm}), 
we proceed as in Ref. \cite{Watanabe:PRA2008,Watanabe:PRL2017} by imposing an order parameter of the Bloch form 
$\Delta(\mathbf r)=e^{i 2 Q x} \tilde \Delta(x) $ in the BDG equations, where 
$\tilde \Delta(x)$ is a periodic function with period $d$. This implies that the corresponding eigenfunctions 
take the form $u(\mathbf r) =
\tilde{u_{\mathbf k}}(x) e^{i Q x}e^{i\mathbf k \cdot \mathbf r }/\sqrt V$ and
$v(\mathbf r) = \tilde{v_\mathbf k}(x) e^{-i Q x}e^{i\mathbf k \cdot
\mathbf r }/\sqrt V$, where $k_x$ lies in the first Brillouin zone 
and
the amplitudes $\tilde{u_\mathbf k}(x)$ and $\tilde{v_\mathbf k}(x)$ are periodic functions with period $d$ obeying 
\begin{equation}
\left( \begin{array}{cc}
H_{Q}(x) & \tilde{\Delta}(x) \\
\tilde{\Delta}^\ast(x) & -H_{-Q}(x) \end{array} \right)
\left( \begin{array}{c} \tilde{u_\mathbf k}(x) \\ \tilde{v_\mathbf k} (x)
\end{array} \right)
=\epsilon_\mathbf k \left( \begin{array}{c} \tilde{u_\mathbf k}(x) \\
\tilde{v_\mathbf k}(x) \end{array} \right) \;,
\label{BdG}
\end{equation}
where
\begin{equation}
  H_{Q}(x)\equiv \frac{1}{2m} \left[\mathbf k_\perp^2
+\left(-i\partial_x+Q+k_x\right)^2\right]+V_{\rm ext}(x) -\mu\, .
\label{hq}
\end{equation}
For a given value of the momentum $\mathbf k$, the BdG equations (\ref{BdG}) represents an eigenvalue problem with the corresponding eigenfunctions
being normalized according to $\int_0^d [\tilde u_{\mathbf k \alpha}^*(x) \tilde u_{\mathbf k \beta}(x)+\tilde v_{\mathbf k \alpha}^*(x) \tilde v_{\mathbf k \beta}(x)]dx=d \delta_{\alpha,\beta}$.

The local paring field $\tilde \Delta (x) $ and the chemical potential 
$\mu$, appearing in Eq.s~(\ref{BdG}) and (\ref{hq}), are variational 
parameters to be determined self-consistently. The first constraint 
is provided by the gap equation
\begin{equation}\label{gap}
\tilde \Delta(x) =- g \frac{1}{V}\sum_{\mathbf k\alpha} \tilde u_{\mathbf k \alpha}(x) \tilde v_{\mathbf k \alpha}^*(x),
\end{equation}
where  $g$ is the bare coupling constant of the contact interaction potential and the sum is restricted over the solutions with 
$\epsilon_{\mathbf k\alpha}\geq 0$.
It is well known that the rhs of Eq.(\ref{gap}) is actually plagued by a ultraviolet divergence and must be regularized. Following the procedure suggested in Ref.\cite{Bulgac:PRL2002},
we introduce an explicit high-energy cut-off in the sum over eigenenergies,  and replace $g$ with a position-dependent effective coupling constant. 
The second constraint to be satisfied is that the spatial average of the density profile must yield the mean density $\bar n$, that is $\bar n=\int_0^d n(x) dx/d$,
where
\begin{equation}\label{n(x)}
 n(x)=\frac{1}{V}\sum_{\mathbf k\alpha} 2    |\tilde v_{\mathbf k \alpha}(x)|^2.
\end{equation}
  
Once the mean field parameters are known, 
we compute the energy density $e=E/V$ of the system according to
\begin{eqnarray}
 e&=& \frac{1}{V}\sum_{\mathbf k\alpha}  \int_0^d \frac{dx}{d}  \Big[2(\mu-\epsilon_{\mathbf k\alpha})  |\tilde v_{\mathbf k \alpha}(x)|^2   \nonumber \\ 
 &&+\tilde \Delta^*(x) \tilde u_{\mathbf k \alpha}(x) \tilde v_{\mathbf k \alpha}^*(x) \Big].   \label{edensity}
\end{eqnarray}
and use Eq.(\ref{invm}) to extract the superfluid fraction from the obtained result. 
Notice that the two contributions in the rhs of Eq.(\ref{edensity}) are separately divergent but their sum is  finite, implying that this equation is well-defined.   

\begin{figure}
\includegraphics[width=\columnwidth]{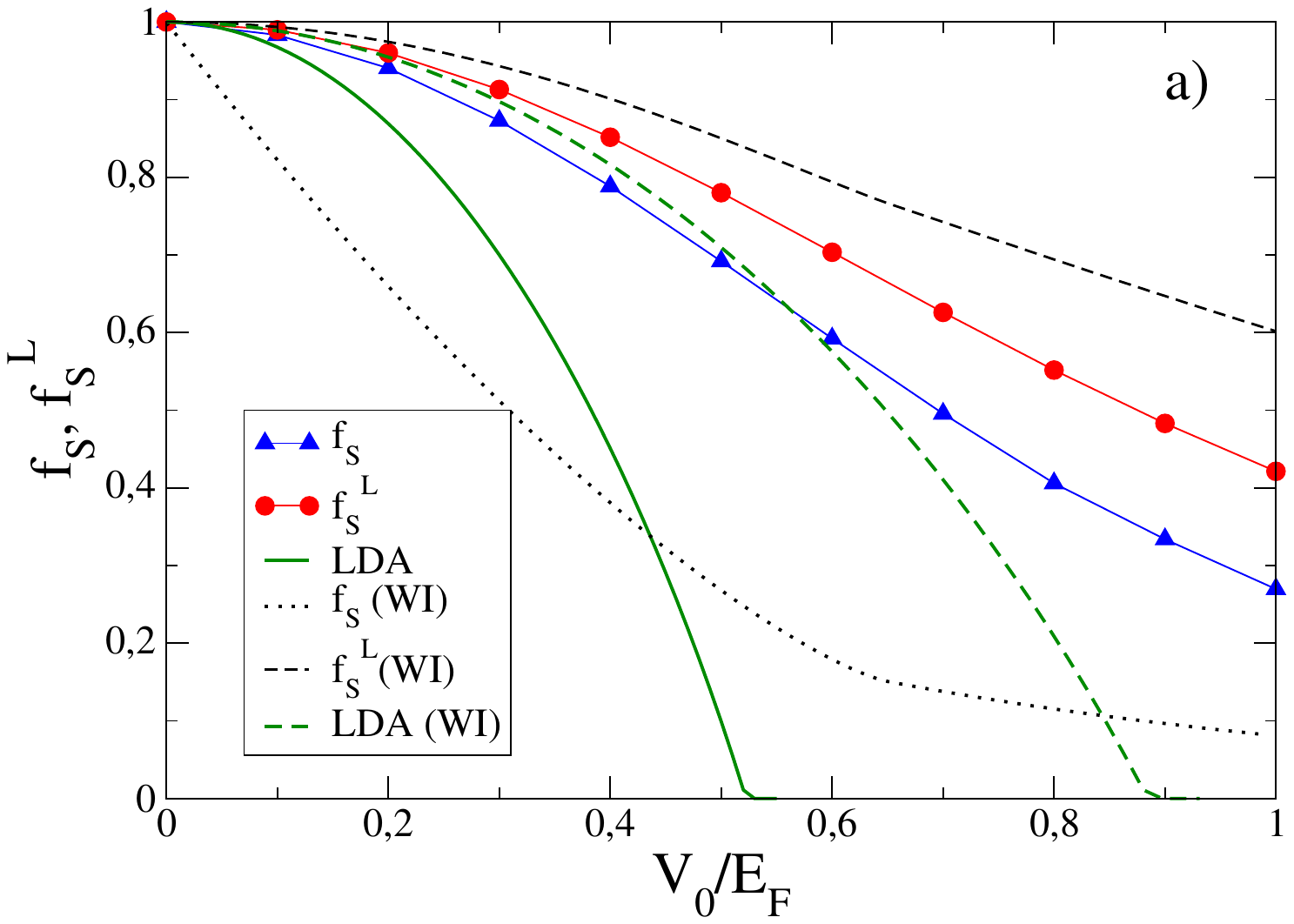}
\includegraphics[width=\columnwidth]{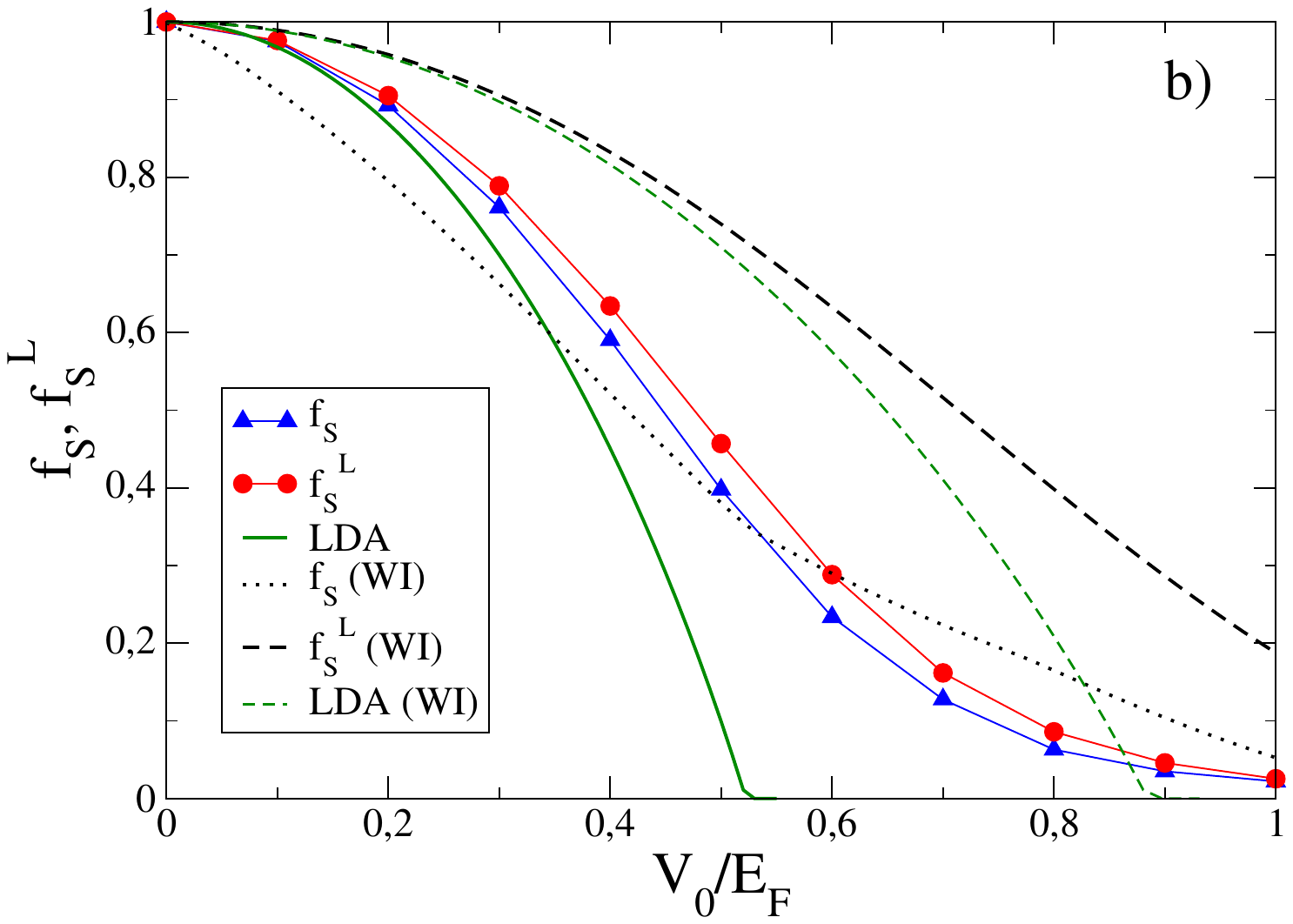}
\caption{Superfluid fraction $f_S$ (blue triangles) and its Leggett's bound $f_S^L$ (red circles) for the unitary Fermi gas as a function of the ratio $V_0/E_F$
for $k_Fd=4$ (upper panel) and $k_Fd=10$ (lower panel). The green solid line refers to the local density approximation (LDA) at unitarity. 
In each plot we also include the corresponding predictions for the weakly interacting BCS regime for $f_S$ (black  dotted line),  $f_{S}^L$ (black dashed line) and for LDA (green dashed line). 
}
\label{Fig:invmass}
\end{figure}

For the Leggett's bound we instead solve the BdG equations (\ref{BdG}) at equilibrium, i.e. with $Q=0$,  and then extract $f_S^L$ from Eq.s (\ref{LB}) and (\ref{n(x)}). In Fig.\ref{Fig:invmass} we display $f_S$ and $f_S^L$ as a function of the ratio $V_0/\epsilon_F$ for the unitary Fermi gas, the two panels corresponding to two 
  different values of the dimensionless parameter $k_Fd=4$ (a) and $k_Fd=10$ (b). 
We see that the Leggett bound is generally not saturated for the unitary Fermi gas,
 but the difference tends to reduce as $k_Fd$ increases. In particular in the limit $k_Fd  \gg 1$ the two values
 $f_S$ and $f_S^L$ are expected to approach the prediction of the local density approximation (LDA), where the density profile of the Fermi gas can be calculated starting from the expression $\mu_0=\mu_\textrm{hom}(n(x))+V_0 \cos(q x)$ for the chemical potential, with $\mu_\textrm{hom}(n)=(1+\beta)E_F=(1+\beta)(3\pi^2 n)^{2/3}/(2m)$ the equation of state of the uniform Fermi gas at unitarity \cite{BecBook2016}.   
 The density profile in the LDA then takes the form
 \begin{equation}\label{LDA}
 n_\textrm{LDA}(x)=\frac{1}{3\pi^2} \left(\frac{2m}{1+\beta}\right)^{3/2} \Big[\mu_0-V_0 \cos(2\pi x/d)\Big]^{3/2} \; ,
 \end{equation}
yielding the prediction 
\begin{equation}\label{fSLDA}
\frac{1}{f_S^{LDA}}=\langle n\rangle \langle \frac{1}{n}\rangle =h\left(\frac{V_0}{\mu_0}\right) \int_0^1 d\tilde x \Big[1-\frac{V_0}{\mu_0}\cos(2\pi \tilde x)\Big]^{-3/2},
\end{equation}
for the inverse superfluid fraction,
 where $\tilde x=x/d$ and the function $h$ is defined as $ h(V_0/\mu_0)=\int_0^1 d\tilde x [1-(V_0/\mu) \cos(2 \pi \tilde x)]^{3/2}$. 
 The chemical potential $\mu_0$ is related to the Fermi energy $E_F$ and the potential strength $V_0$ via the normalization condition $\bar n=\int_0^d n_\textrm{LDA}(x) dx/d$, yielding  
\begin{equation}\label{mu0}
\mu_0=\frac{(1+\beta)E_F}{ h(V_0/\mu_0)^{2/3}}.   
\end{equation}
Eq.s (\ref{fSLDA}) and (\ref{LDA}) show that in the  LDA the superfluid density  is independent of the lattice period $d$ and only depends on the ratio 
$V_0/E_F$ (see green solid line in Fig.\ref{Fig:invmass}). In particular the superfluid fraction vanishes for $V_0/E_F= (1+\beta)/h(1)^{2/3}=0.522$, corresponding to $V_0/\mu_0=1$, where
the integral in Eq.(\ref{fSLDA}) diverges.  

For small values of $V_0/E_F$ the LDA expression for the superfluid density exhibits the quadratic dependence (see also \cite{Watanabe:PRA2008} and Eq.(\ref{fsm-3}) in the $q\to 0$ limit)
\begin{equation}
f_S^{LDA}= 1 - \frac{9}{8 (1+\beta)^2} \frac{V_0^2}{E_F^2},
\label{LDAlowq}
\end{equation}
It is interesting to compare the above expansion  with the 
corresponding expansion for the compressibility, in the same 
LDA regime ($q\to 0$ limit): 
\begin{equation}
\kappa^{LDA}= \frac{2}{3} (1+\beta)E_F \Big [1 + \frac{1}{8 (1+\beta)^2} \frac{V_0^2}{E_F^2} \Big],
\label{kappalowq}
\end{equation}
 Eq.(\ref{kappalowq}) is easily obtained by applying  perturbation theory for the calculation of the energy in the presence of the perturbation (\ref{Vext}) and plays a crucial role in the propagation of sound (see Sect IV below). 
Notice in particular that the effect of the perturbation is  smaller  by a factor $9$, with respect to the analogous expansion (\ref{LDAlowq})
for the superfluid fraction. It is also interesting to compare the above results with the expansions holding in a dilute BEC gas \cite{Chauveau:PRL2023} where, in the LDA the $V_0^2$ correction to the compressibility is exactly vanishing, as a consequence of the linear density dependence exhibited by the equation of state.

In Fig.\ref{Fig:invmass} we also show the superfluid fraction (dotted line) and its Leggett's bound (dashed line) for a Fermi gas in the weakly interacting (WI) BCS regime,  corresponding to $\Delta \ll E_F$.  In this limit the density profile and the energy of the system can be  calculated by assuming that the Fermi gas is non interacting.
 Of course the non interacting Fermi gas  is not superfluid because of the occurrence of transverse gapless excitations. However,  in the superfluid phase the superfluid fraction  $f_S$ coincides with the effective mass  (see (Eq.(\ref{invm})). The  evaluation of the effective mass in the non interacting Fermi gas is expected to provide the actual value of $f_S^{WI}$, if the condition $qk_F/m\gg \Delta$, corresponding to wave lengths of the perturbation much smaller than the size of Cooper pairs,  is satisfied.
 One can then write \cite{Pitaevskii:PRA2005}:
\begin{equation}\label{fSNI}
f_S^{WI}\simeq \frac{1}{\bar nV}  \sum_{\mathbf k j} \frac{\partial^2 \epsilon_j (k_x) }{\partial k_x^2} 2 \Theta\Big(\mu-\epsilon_j (k_x) -\frac{\mathbf k_\perp^2}{2m}\Big),
\end{equation}
where $\epsilon_j (k_x)$ are the single-particle dispersion relations induced by the 1D  external potential, with $j$ being the band index, and the integration over $k_x$ is restricted to the first Brillouin zone. 

The Leggett bound in the  WI regime can be obtained from Eq.(\ref{LB}) by using the density profile of the non interacting Fermi 
gas 
\begin{equation}
   n(x)=\frac{1}{\bar nV}  \sum_{\mathbf k j} |\psi_{k_x j}(x)|^2 2 \Theta\Big(\mu-\epsilon_j (k_x) -\frac{\mathbf k_\perp^2}{2m}\Big), 
\end{equation}
where $\psi_{k_x j}(x)$ are the amplitudes of the Bloch states.

A first useful comparison between the superfluid fractions calculated at unitarity and in the weekly interacting BCS gas concerns the behavior of  Leggett bound in the local density approximation (see Eq.(\ref{fSLDA})). Since  the chemical potential of the non interacting Fermi gas and of the gas at unitarity have  the same density dependence, except for the factor $(1+\beta)$ (see Eq.(\ref{mu0})), one finds that in the weakly  interacting case $f_S^{LDA}$ always stays significantly {\it{above}} the corresponding value at unitarity. In particular the LDA superfluid fraction  vanishes for a larger value of the ratio $V_0/E_F$, equal to $0.522/(1+\beta)=0.883$. 
The situation can be however very different for the actual value $f_S$ of the superfluid fraction. For example, for small values of $V_0$ the superfluid fraction of the weakly interacting BCS gas always stays  {\it{below}} the value at unitarity. 
For large wave vectors (see Fig. \ref{Fig:invmass}a) the same happens  also for larger values of $V_0$, suggesting that 
interactions play a crucial role to facilitate superfluid transport, as previously reported by Watanabe and Pethick with important implications on the behavior of  neutrons stars \cite{Watanabe:PRL2017}.
 
Another interesting feature emerging from  Fig.\ref{Fig:invmass} (ab) is that  in the WI regime $f_S$ exhibits a linear dependence for small $V_0$. We have checked that this behavior occurs only for $k_Fd\geq \pi$, while for $k_Fd<\pi$ we  find that $1-f_S \propto V_0^2$, as previously observed for the interacting system. This effect can be simply understood by noting that for a noninteracting uniform Fermi gas the moment $m_{-3}(q)$ in Eq.(\ref{fsm-3})  diverges for $q<2k_F$, because the dynamic structure factor $S(q,\omega)$ is linear in $\omega$  at low frequency, reflecting the presence of gapless single-particle excitations at the Fermi surface. In the superfluid  Fermi gas these  excitations are instead gapped due to Cooper pairing. 
In the Appendix we provide an explicit derivation of the behavior of $f_S$ for small $V_0$ in the WI regime based on degenerate perturbation theory.

\section{IV. Behaviour of  the sound velocities}

The results discussed in the previous Section on the superfluid density of the interacting Fermi gas have important implications on the value of the sound velocity and in particular on its anisotropy in the presence of the periodic perturbation (\ref{Vext}).
According to the hydrodynamic theory of superfluids, the square sound velocities  along the longitudinal and transverse  directions are given by
\begin{eqnarray}
\label{cxcy}
 c^2_x&=& f_S\kappa^{-1}\nonumber \\ 
 c^2_y&=&c^2_z= \kappa^{-1},
\end{eqnarray} 
emphasizing that their anisotropy,  caused by the presence of the 1D periodic potential, follows from the anisotropy of the superfluid fraction.   
From an experimental point of view a natural strategy to determine the value of the superfluid fraction, associated with the motion of the superfluid along the $x$-direction, then consists of measuring the ratio 
\begin{equation}
 \frac{c^2_x}{c^2_y}= f_S   \; 
\end{equation}
which is independent on  the value of the compressibility.  
This procedure has been recently successfully employed in the case of a dilute Bose gas confined in a 2D box trap \cite{Chauveau:PRL2023}.

\section{V. CONCLUSIONS}
In this work we have investigated the superfluid fraction of a Fermi gas at unitarity, in the presence of a one-dimensional periodic potential. We have shown that, in sharp contrast with the dilute BEC gas case,  the Leggett's upper bound for the superfluid fraction can significantly overestimate its actual value at zero temperature, unless the period $2\pi/q$ of the applied lattice becomes much larger than the size of Cooper pairs, 
i.e. if  $q\ll \Delta/(mk_F)$, where $\Delta$ is the pairing gap energy. We have pointed out the occurrence of important deviations of Leggett's bound from the actual value $f_S$ of the superfluid fraction  at unitarity and shown that these deviations becomes more and more important as one approaches the weakly interacting BCS regime of small and negative values of the $s$-wave scattering length. Using sum rule arguments we have also shown that the deviations of the Leggett bound from the superfluid fraction explain the change of sign of the curvature of the Anderson-Bogoliubov mode dispersion taking place around unitarity. 

We thank  Y. Castin, F. Dalfovo, S. Giorgini, A. Leggett and W. Zwerger for fruitful discussions.
This work was granted access to the HPC resources of TGCC under the allocations AD010513635 and AD010513635R1 supplied by GENCI (Grand Equipement National de Calcul Intensif). S.S. acknowledges continuous financial support from PAT (Provincia Autonoma di Trento.)

\section{APPENDIX}
For completeness, in this Appendix we provide an analytical calculation of $f_S$ for a weak periodic potential 
in the weakly interacting BCS regime, under the assumption
that $q\gg \Delta m/k_F$, which is equivalent to $k_F d\ll E_F/\Delta$. As discussed in the main text, under this condition one is allowed to identify the inverse effective mass, calculated using the ideal Fermi gas, with the superfluid fraction of the weakling interacting BCS superfluid. Below we will consider separately two cases, $k_F d <\pi$ and $k_F d\geq \pi$. 

\subsection{Case $k_Fd<\pi$}
We calculate $f_S^{WI}$ from Eq.(\ref{fsm-3}) since $m_{-3}(q)$ is expected to be finite for a uniform system. For $q>2k_F$  the dynamic structure factor of the noninteracting Fermi gas is given by
\begin{equation}\label{SqNI}
 S(\omega,q)=\frac{m}{4\pi^2 q}\left[k_F^2-\frac{m^2}{q^2} \left (\omega-\frac{q^2}{2m}\right)^2\right],
\end{equation}
if the rhs of Eq. (\ref{SqNI}) is positive, that is for $|\omega-q^2/(2m)|<qk_F/m$, and zero otherwise.
\begin{eqnarray}
m_{-3}(q)&=&\frac{m}{4\pi^2 q \bar n} \int_{q^2/(2m)-q k_F/m}^{q^2/(2m)+q k_F/m} \frac{d\omega}{\omega^3} \nonumber\\
&&\times \left[ k_F^2-\frac{m^2}{q^2} \left (\omega-\frac{q^2}{2m}\right)^2   \right]   
\end{eqnarray}
To perform the integral, we introduce dimensionless variable $y=[\omega-q^2/(2m)]/E_F$ and $p=q/k_F$, so that the integral becomes
\begin{equation}\label{A1}
 m_{-3}(q)=\frac{m^3}{\pi^2 k_F^3 \bar n p} 
 \int_{-2p}^{2p} 
 \left(1-\frac{y^2}{4 p^2}\right)\frac{dy}{(y+p^2)^3}   
\end{equation}
By performing the integration in Eq.(\ref{A1}) and recalling that $\bar n=k_F^3/(3\pi^2)$, we obtain  $m_{-3}(q)= F(q/k_F)/E_F^3$, where
\begin{equation}\label{Fp}
F(p)=\frac{3}{8 p^3}\left[\ln\left(\frac{p-2}{p+2} \right)+\frac{4p}{p^2-4}\right].
\end{equation}
 Since $m_1(q)=q^2/(2m)$, from Eq.(\ref{fsm-3}) we find 
\begin{equation}
1-f_S^{WI}\simeq 2 \left (\frac{q}{k_F}\right)^2F\left(\frac{q}{k_F}\right) \left (\frac{V_0}{E_F}\right)^2.  
\end{equation}

Notice that the function $F(p)$ in Eq.(\ref{Fp}) diverges at $p=2$ as $F(p)\approx \frac{3}{128 (p-2)} $, implying that for $q=2k_F$ the correction to the superfluid fraction induced by the periodic potential are no longer quadratic but linear in $V_0$ ,  as we shall see below. 

\subsection{Case $k_Fd\geq \pi$ }
We use degenerate perturbation theory applied to the external periodic potential (\ref{Vext}) to show that $1-f_S\propto V_0$. For simplicity, we will prove the claim for the special case $k_Fd=\pi$, so that only the lowest energy band is populated. The proof can be easily adapted to the general case where higher bands are also populated. 

We consider two states with wave-vector
$k_x$ and $k_x-G$, where $G$ is a reciprocal lattice point. The two energy bands are  obtained by the condition  
\begin{equation}
\begin{vmatrix}
\frac{k_x^2}{2m}-E & \frac{V_0}{2} \\ 
\frac{V_0}{2} & \frac{(k_x-G)^2}{2m}-E
\end{vmatrix}
=0,
\end{equation}
whose solutions are 
\begin{eqnarray}\label{Ept}
    E_\pm (k_x) &=&  \frac{1}{2} \left(   \frac{k_x^2}{2m}+\frac{(k_x-G)^2}{2m}\right)\nonumber \\
&&\pm \left [\frac{1}{4}  \left(   \frac{k_x^2}{2m}-\frac{(k_x-G)^2}{2m}\right)^2+\frac{V_0^2}{4} \right]^{1/2}.
\end{eqnarray}
We consider $0\leq k_x\leq \pi$ and choose $G=2\pi/d$ in Eq.(\ref{Ept}) so that  $E_\pm (\pi/d)=E_R\pm V_0/2$. For $k_F=\pi/d$, the Fermi energy $E_F=E_R$ sits between the two bands, so the upper band is completely empty and according to in Eq.(\ref{fSNI}) it does not contribute to the superfluid fraction.
Let us introduce the dimensionless variables $\tilde E_-=E_-/E_R$ and $x=k_x \pi/d$. Then
\begin{equation}\label{Eminus}
   \tilde E_-(x)=x^2+2-2x-[4(1-x)^2+V_r^2]^{1/2}, 
\end{equation}
where $V_r=V_0/(2E_R)$ is the rescaled potential strength. From Eq. (\ref{Eminus}) we find
\begin{equation}\label{curv}
\frac{\partial^2 \tilde E_{-}(x)}{\partial x^2} =
2-\frac{4V_r^2}{[4(1-x)^2+V_r^2]^{3/2}},
\end{equation}
showing that the curvature of the bands diverges as $2-4/V_r$ for vanishing $V_r$.  
Substituting Eq.s (\ref{Eminus}) and (\ref{curv}) into Eq.(\ref{fSNI}), we find 
\begin{equation}\label{partialfSNI}
f_S^{WI} \simeq 1- \frac{4 V_r^2 \int_0^1 \frac{[-(x-1)^2+\sqrt{4(1-x)^2+V_r^2}]}{[4(1-x)^2+V_r^2]^{3/2}} dx}{2\int_0^1 [-(x-1)^2+\sqrt{4(1-x)^2+V_r^2}]dx}.
\end{equation}
Let us discuss the behavior of the two integrals in Eq.(\ref{partialfSNI}), starting from the upper one. Since the major contribution to the integral comes from the region around $x=1$, the term proportional to $-(x-1)^2$
in the numerator can be neglected as it vanishes for $x=1$ (its contribution to the integral is of order $V_r^2 \ln V_r$). For the lower integral, we can safely set $V_r=0 $ as no singularity is present, so that Eq. (\ref{partialfSNI}) reduces to  
\begin{equation}
f_S^{WI} \simeq  1- \frac{4 V_r^2 \int_{0}^1 [4(1-x)^2+V_r^2]^{-1}dx}
{2 \int_0^1 [-(x-1)^2+ 2(1-x) ]dx}.   
\end{equation}
Since $V_r $ is small, in the upper integral we can safely replace $0$ with $-\infty$, so that the integration  becomes elementary, yielding
\begin{equation}
 f_S^{WI} \simeq 1-\frac{\pi V_r}{4/3}  =1-\frac{3\pi}{8}\frac{V_0}{E_F}, 
\end{equation}
showing that the correction to the superfluid fraction is linear in $V_0$, which  agrees with our numerics.

\bibliography{bibliosupersolid.bib}

\begin{thebibliography}{23}
\bibitem [{not()}]{notefS}%
  \BibitemOpen
  \href@noop {} {\bibinfo {title} {In general the superfluid density (and hence
  the superfluid fraction) is an anisotropic tensor and in the following we
  will adopt the notation $f_s$ to indicate the superfluid fraction along the
  $x$-axis.}}\BibitemShut {Stop}%
\end{thebibliography}

\end{document}